%% file: main.tex
\documentclass[9pt,letter]{sigcomm-alternate}

\usepackage{epstopdf}
\usepackage{xspace,epsfig}
\usepackage{graphicx,times,latexsym,amsfonts,amssymb,amsmath}
\usepackage{comment,url,color}
\usepackage{multirow}
\usepackage{caption}
\usepackage{subcaption}
\usepackage{enumitem}

\newcommand{\diff}{{\rm\,d}}

\newcommand{\ls}[1]
{\dimen0=\fontdimen6\the\font
  \lineskip=#1\dimen0
  \advance\lineskip.5\fontdimen5\the\font
  \advance\lineskip-\dimen0
  \lineskiplimit=.9\lineskip
  \baselineskip=\lineskip
  \advance\baselineskip\dimen0
  \normallineskip\lineskip
  \normallineskiplimit\lineskiplimit
  \normalbaselineskip\baselineskip
  \ignorespaces
}

\newcommand{\tgifeps}[3]{
  \begin{figure}[t]
    \centering
    \includegraphics[width=#1.0cm, clip=true]{figures/#2.eps}
    \caption{#3\label{fig:#2}}
    \vspace{-3.5mm}
  \end{figure}

}

\includecomment{conference}
\excludecomment{techrep}

\newcommand{\be}{\begin{equation}}
\newcommand{\ee}{\end{equation}}
\newcommand{\ba}{\begin{array}}
\newcommand{\ea}{\end{array}}

\newcommand{\youtube}{{YouTube}\xspace}

\newcommand{\RMMOA}[2]{}

\newcommand{\TONE}{\textsc{Trace 1}\xspace}
\newcommand{\TTWO}{\textsc{Trace 2}\xspace}
\newcommand{\TTHREE}{\textsc{Trace 3}\xspace}
\newcommand{\TFOUR}{\textsc{Trace 4}\xspace}
\newcommand{\TALL}{\textsc{Traces 1-4}\xspace}

\begin{document}
\begin{sloppypar}

\title{Temporal Locality in Today's Content Caching:  Why it Matters and How to Model it}


\numberofauthors{3}
\author{
  \alignauthor
  Stefano Traverso \\
  \affaddr{Politecnico di Torino}\\
  \affaddr{Torino, Italy}\\
  \email{stefano.traverso@polito.it}
  \alignauthor
  Mohamed Ahmed \\
  \affaddr{NEC Labs Europe}\\
  \affaddr{Heidelberg, Germany}\\
  \email{mohamed.ahmed@neclab.eu}
  \alignauthor
  Michele Garetto \\
  \affaddr{Universit\char224\  di Torino}\\
  \affaddr{Torino, Italy}\\
  \email{\ \ michele.garetto@unito.it}
  \and
  \alignauthor
  Paolo Giaccone \\
  \affaddr{Politecnico di Torino}\\
  \affaddr{Torino, Italy}\\
  \email{paolo.giaccone@polito.it}
  \alignauthor
  Emilio Leonardi \\
  \affaddr{Politecnico di Torino}\\
  \affaddr{Torino, Italy}\\
  \email{emilio.leonardi@polito.it}
  \alignauthor
  Saverio Niccolini \\
  \affaddr{NEC Labs Europe}\\
  \affaddr{Heidelberg, Germany}\\
  \email{saverio.niccolini@neclab.eu}
  }
\maketitle

\input{abstract}

\vspace{1em}
 \noindent {\bf  Categories and Subject Descriptors} 
C.2.1 [{ Network Architecture and Design}]

\noindent{\bf General Terms}: 
 Theory, Measurement and Modelling


\input{intro}

\input{necessity2}
\input{analysis3}

\input{traffic2}


{
\bibliographystyle{abbrv_ccr}
\bibliography{CCR}
}

\begin{techrep}
\appendices
\input{appendix}
\end{techrep}
\end{sloppypar}

\end{document}

%% file: abstract.tex
\begin{abstract}

  The dimensioning of caching systems represents a difficult task in
  the design of infrastructures for content distribution in the
  current Internet. This paper addresses the problem of defining a
  realistic arrival process for the content requests generated by
  users, due its critical importance for both analytical and
  simulative evaluations of the performance of caching systems.
  First, with the aid of \youtube traces collected inside operational
  residential networks, we identify the characteristics of real
  traffic that need to be considered or can be safely neglected in
  order to accurately predict the performance of a cache.  Second, we
  propose a new parsimonious traffic model, named the Shot Noise Model
  (SNM), that enables users to natively capture the dynamics of
  content popularity, whilst still being sufficiently simple to be
  employed effectively for both analytical and scalable simulative
  studies of caching systems.  Finally, our results show that the SNM
  presents a much better solution to account for the temporal locality
  observed in real traffic compared to existing approaches. 

\end{abstract}

%% file: intro.tex
\section{Introduction}
\label{sec:introduction}
Content caching plays a critical role when operating networks by
reducing the distance packets travel in the network. This results in
lowered costs for operators and improved quality of service for end
users.  It is therefore no surprise that content caching plays a
central role in many recent approaches aimed at improving network
performance and utilisation, including; Information Centric Networking
(ICN)~\cite{Jacobson_icn,Roberts_mix}, resource management in cellular
networks~\cite{Erman-GSM,Qian_2012}, energy efficient
networking~\cite{Lee:2011}, massively geographically distributed
CDNs~\cite{Jiang_conext12} and ISP-assisted CDNs~\cite{Poese:2012}.

However, when evaluating the benefits of their proposals, researchers
are faced with the familiar question of what to evaluate their
methodology on. This is particularly poignant for networks researchers
looking to understand the impact of millions of users accessing many
millions of content objects on a network.  The usual methodology here
is to perform trace-driven analysis, collecting traces of the
phenomena in question and using these to drive simulations and
emulations~\cite{Erman-GSM, Qian_2012, Jiang_conext12, Poese:2012,
  Erman:2009}.

There are however two issues here. First, trace-driven analysis is
effective only when large data sets are available, which, for most
researchers is not always the case. Too often we are limited by the
size and/or the availability of data sets, their diversity, or legal
and privacy concerns. Second, this kind of analysis does not allow
users to test potential changes in the traffic profile, such as
changes in the popularity profiles of contents before they actually
take place. In short, in order to evaluate the performance of caching
systems, we must build models that enable us to evaluate effects that
are too large to test in the wild, or for which there are no data
traces available, or that correspond to scenarios which do not yet
exist in operational systems.

In this work we focus on the problem of defining a model to accurately
approximate content request rates, in particular for Video-on-Demand
(VoD) systems. This problem is at the heart of determining the
accuracy of the many previously published simulation and model-based
results that aim to understand the performance of caching
systems. This is important because given the preponderance of
multimedia traffic in today's networks~\cite{Erman:2009}, it is
reasonable to assume that the performance of the network as a whole is
largely dominated by the effects of content delivery, and will become
more so in the future.


Our contribution in this space is two-fold. First, with the aid of
real traffic traces, we present the first quantification of the cache
system performance errors introduced by the classical {\em Independent
  Reference Model} (IRM)~\cite{Coffman:73} (the de-facto standard
synthetic traffic model~\cite{Roberts_CHE, ager_infocom2010}) in the
context of on-demand video delivery. We show that the IRM leads to
considerable errors, rendering its use in this context misguided at
best. We believe this to be an important result, because, due its
tractability, the IRM is still the standard assumption made in many
works~\cite{Lee:2011, Jiang_conext12, Laoutaris2007, irm04}. As our
second contribution, we propose a novel replacement to the IRM called
the {\em Shot-Noise Model} (SNM). The SNM overcomes the IRM's
limitations by explicitly accounting for the temporal locality in
requests for contents, while still maintaining its desirable properties
of simplicity and scalability.  Finally, we validate the SNM's
accuracy using real traffic traces.

The temporal locality in content requests, and its potential effects
on caching systems have been previously studied by a number of works,
including \cite{Almeida:1996, Jin00sourcesand, Fonseca:03,jelekovic}.
However, these works tended to consider early WWW traffic whose
profile differs significantly from today's Video-on-Demand traffic. As
such, previously proposed models tended to laregly ignore the
long-term popularity evolution of contents, in favour of capturing the
short-time scale correlations of the contents request processes.
Moreover, previous works have primarily focused on characterising the
distribution of inter-requests times for a given
content~\cite{Jin00sourcesand, Fonseca:03}, or on the distribution of
content requests that fall between two consecutive requests for a
given content~\cite{Almeida:1996}, providing only a partial
characterisation of temporal locality. Further, in contrast to this
work, previous works have not proposed a phenomenological model that
captures the fundamental origins of temporal locality.

%% file: necessity2.tex
\section{The standard approach}
\label{sec:necessity}
The most common approach to evaluating the performance of caching
systems is to assume that content requests are generated under the
IRM, with a request distribution following a generalised Zipf
law~\cite{Jiang_conext12, Roberts_CHE, ager_infocom2010,
  Laoutaris2007,kurose2013}. The IRM considers a fixed population of
$N$ contents, such that the sequence of content requests arriving at
the cache is characterised by the following set of fundamental
properties: the probability of a request for a given content $n$, for
$1 \leq n \leq N$, is constant (i.e.\ the content's popularity does
not vary over time) and independent of all past requests. Finally, the
set of available contents does not change over time. While in its
simplest form, Zipf's law states that the probability of requesting
the $n$th most popular item is proportional to $1/n^\alpha$, where the
exponent $\alpha$ depends on the considered system (especially on the
type of contents)~\cite{Roberts_mix}.

The IRM is widely adopted because of its simplicity and effectiveness
in facilitating the development of tractable analytic models under
various caching policies~\cite{jel08,dan90,che02}. Although clearly,
the assumptions made by the IRM are too rigid for real traffic, it has
been advocated as an acceptable approximation, when, the popularity
variations of contents over time are relatively slow, with respect to
the time-scale of the content churn in the cache~\cite{Roberts_CHE}.
Whilst extensions of the IRM, aimed at capturing temporal correlations
in the request process have been proposed in the literature, these
tend to focus on applications other than
video-on-demand~\cite{coffman73, virtamo98, jelekovic}.  Further, in
contrast to the SNM, these works assume that the requests for each
content are drawn from a fixed catalogue, and follow a stationary
process (either a renewal process, or a Markov or Semi-Markov
Modulated Poisson process). They therefore ignore the temporal
evolution of content popularity, which plays a crucial role in the
context of on-demand video.

In the rest of this section, we assess the applicability of the
arguments for the IRM by critically discussing the common assumptions
made when utilising it. Using real data traces, we show that when
applied to cache performance analysis, the IRM assumptions lead to
significant errors.

\input{dataset}

\subsection{Mistakes and Limitations of the IRM}
\label{sec:mistakes}
The main difficulty that is encountered when adopting an IRM approach
is the specification of the content popularity distribution, subject
to a given content catalogue of size $N$. This distribution must be
chosen with extreme care, since the cache performance depends heavily
on the relative popularity of different contents~\cite{Roberts_mix,
  Jin00sourcesand}.  While the choice of a Zipf distribution (and its
generalisations) is supported by experimental evidence, including
measurements of web pages, file sharing, video-on-demand and user
generated content access profiles, its parameterisation is not
entirely obvious.

\tgifeps{8}{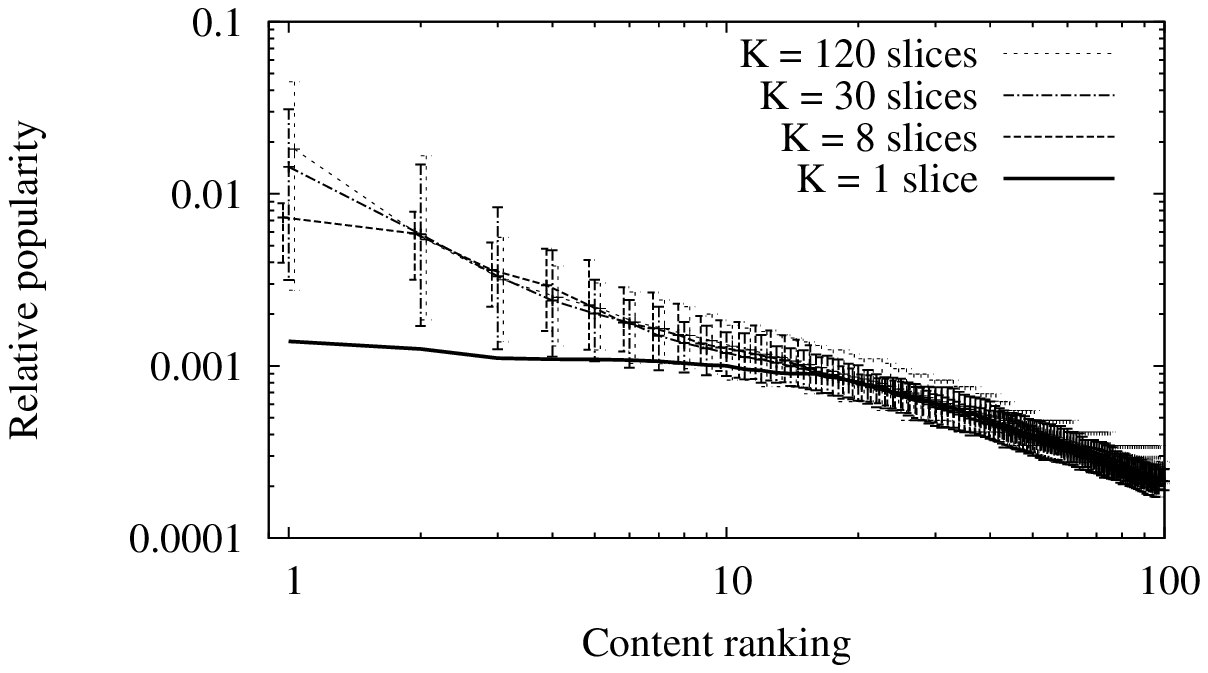}{The empirical popularity distributions of the 100
  most popular videos observed in \TONE. Each distribution is
  generated by splitting the trace into $K$ slices, where ``K=1''
  corresponds to the entire trace. For each $K$, we report the average
  of the relative frequency of each rank position across all the
  slices, and the error bars correspond to 5-95 percentiles.}

\tgifeps{8}{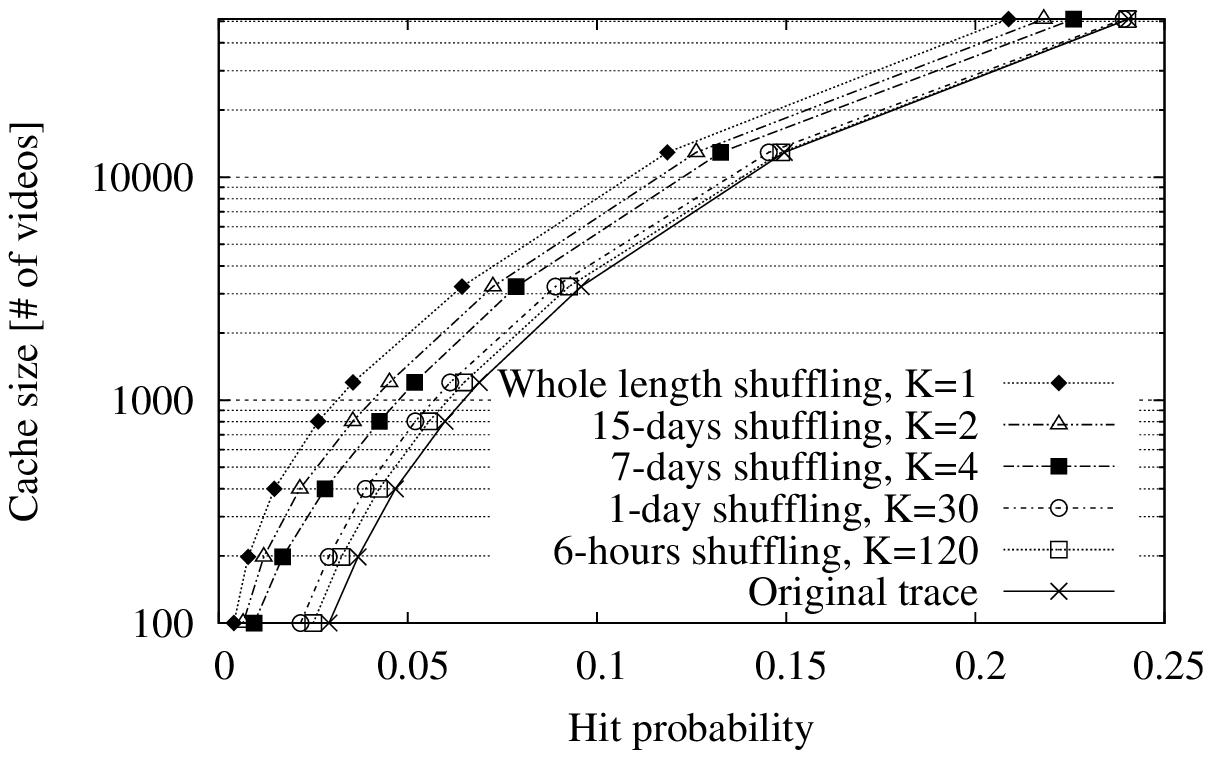}{The cache size
  required to achieve a desired hit probability, when an LRU cache is
  populated by the requests contained in \TONE, subject to differing
  degrees of trace reshuffling. Note that $K=1$ corresponds to the
  basic IRM.}

In fact, when determining a representative content request rate, given
the availability of a long data trace, it is easy for practitioners to
fall into the trap of directly plugging into the cache model, a
popularity distribution estimated from long-term measurements. Even
when it is reasonable to assume that cache dynamics are much faster
than popularity variations (so that we can apply a time-scale
separation), the instantaneous popularity distribution can differ
significantly from the distribution derived from long-term measurement
campaigns.

We use \TONE to illustrate the impact of directly using the popularity
distributions estimated from long-term measurements. To do so, we
derive different empirical popularity distributions for the contents
in the trace by first dividing it into $K$ slices, such that each
slice contains exactly the same number of requests. We then compute
the relative popularity of the contents in each slice, with respect to
the total number of requests observed in the slice (normalised to
one).  The results of this experiment are reported in
Fig.~\ref{fig:fig1}, in which we plots, for each rank position of the
100 most popular videos, the average and the $5-95$ percentiles for
the relative frequency of requests (evaluated across the $K$ slices).

From Fig.~\ref{fig:fig1}, we observe that the steepness of the
empirical distribution for the content popularity increases with
respect to the number of slices considered. In practice, increasing
$K$ corresponds to having larger estimates of $\alpha$ for the Zipf
distribution. Fitting the value of $\alpha$ for the tail of the
distributions, we find that $\alpha$ ranges from $0.70$ ($K=1$) to
$0.85$ ($K=120$). Moreover, as shown by the widening error bars,
increasing $K$ results in a higher variance in the relative popularity
of the contents. The practical consequence of this observation is that
estimates of the popularity distribution become more noisy as we
decrease the time scale we consider (increasing K) . In short, it
becomes increasingly more difficult to derive from the trace a
reliable estimate of the popularity distribution as we consider
decreasing time scales.

The result is that the error arising from using the long-term
popularity distribution of contents, to predict cache performance
(following the IRM approach), can be significant for a cache
implementing the Least Recently Used (LRU) eviction policy
(see Fig.~\ref{fig:phit_vs_cachesize_shuffle_flip_trace_01}).  The
reason for this can be understood when we consider that the above
error is equivalent to that introduced by completely neglecting all
temporal correlations (the so-called ``temporal locality'') in the
arrival process of requests.  To see this, suppose that we shuffle at
random a trace of requests, the result would be that the long-term
popularity distribution resulting from the shuffled sequence remains
exactly the same, but all temporal correlations would be broken.

We could still pursue the IRM approach and correct for the loss in
temporal correlation, however, this is not a straightforward task. To
do so, we would first need to determine an ``evaluation interval'',
that is comparable with the time-scale of cache dynamics, for the
content popularity. Assuming that the popularity variations over the
interval are negligible, we would then need to compute a modified
popularity distribution for the contents requested in the
interval. Notice that in doing so, we would need to compute it on a
reduced catalogue of size much less than $N$, which is necessary to
properly normalise the request probabilities relative to the interval.

Finally, to show the impact of the loss in temporal locality when
evaluating the performance of a cache, in
Fig.~\ref{fig:phit_vs_cachesize_shuffle_flip_trace_01}, we compare the
cache size required to achieve a given cache hit rate, when (i) using
the original trace (\TONE), and (ii) synthetic traces derived from the
trace in which we control the degree of temporal locality in the
trace. To construct the synthetic traces, we again partition the
original trace into $K$ slices, then randomly permute the requests
within each slice to remove their temporal locality, such that,
consecutive requests become independent.

It is worth noting that due to the limited duration of our traces, we
are constrained to considering cache sizes such that the eviction time
(i.e., the time since the last request after which a content is
evicted from the cache) is significantly shorter than the duration of
the whole trace.  Under the request rates observed in our traces, the
average eviction time is 2-3 days for cache size in the order of
$50,000$ objects, which is the maximum considered in our
experiments. This explains why in
Fig.~\ref{fig:phit_vs_cachesize_shuffle_flip_trace_01} we could only
report hit probabilities in the range [0, 0.25].

From Fig.~\ref{fig:phit_vs_cachesize_shuffle_flip_trace_01}, the case
where $K=1$, corresponds to breaking all temporal correlations (i.e.,
the result of a naive application of the IRM approach). For instance,
the sequence 11...1, 22...2, 33...3, which is clearly not
independent would succeed most tests of independent sequence after
shuffling with $K=1$. By increasing $K$, we reduce the average size of
the time window considered by each slice and obtain artificial traces
that are increasingly more similar to the original trace, i.e.,
temporal correlations are broken only within slices of diminishing
length. Fig.~\ref{fig:phit_vs_cachesize_shuffle_flip_trace_01} shows
two important findings: first, the IRM assumption leads to a
significant over-estimation (by a factor between 2 and 10) of the
required cache size, specially when the hit probability is low.
Second, the impact of increasing $K$ (i.e., decreasing the temporal
duration of slices) is that the required cache size approaches the
result for the unmodified trace, especially as the slice duration
approaches the order of a few hours.

The results in Fig.~\ref{fig:phit_vs_cachesize_shuffle_flip_trace_01}
suggest that in this particular scenario, the IRM approach could
produce accurate predictions of cache performance, provided that we
are able to estimate the relative popularity of contents requested
within intervals with a duration of around one day.  However, this
approach has several drawbacks:

\begin{itemize}[leftmargin=*,nolistsep,noitemsep,topsep=0pt,parsep=0pt,partopsep=0pt]
\item As shown in Fig.~\ref{fig:fig1}, due to the random fluctuations
  and possible long-range effects in the trace, it is in practice very
  hard to obtain from measurements, a content popularity distribution
  for short time-scales.
\item It is even harder to check that popularity variations are
  negligible over the chosen interval, especially for the least
  popular contents, for which we will have fewer samples.
\item In practice, the evaluation interval should be adapted to the
  time-scale of cache dynamics. This inevitably depends on several
  factors (aggregate requests arrival rate, cache size, caching
  policy, etc.), and forces users to recompute a different popularity
  distribution for each scenario.
\item The time-scale of cache dynamics itself depends on the
  popularity distribution that we are trying to characterise, leading
  to a circular dependency. 
\item It is even possible (although only for very large caches and/or
  small request rates) that popularity variations cannot be considered
  negligible with respect to cache dynamics, rendering the IRM approach
  inaccurate.
\end{itemize}

With respect to the listed reasons, and in contrast to the common
opinion, we believe that the IRM approach has severe limitations,
especially in scenarios characterised by dynamic contents with
evolving popularity profiles.  The consequence is therefore a need for
alternative approaches capable of recapturing the temporal locality in
content requests. Before presenting our solution, we first take a
closer look at the behaviour of content requests in our traffic
traces.

%% file: dataset.tex
\subsection{Data Set}
\label{sec:dataset}

\begin{table}
\centering
\scriptsize
\caption{The experiment traces.}
\begin{tabular}{|c|c|c|c|c|c|c|c|}\hline
Trace 		& PoP		& Period (2012)    & Length & IPs	& Requests & Videos	\cr\hline
\TONE\ 		& PoP 1	& 20/03-25/04 	& 35 days	& 14224		& 1.7M	   & 0.93M\cr
\TTWO\		& PoP 1   	& 30/04-28/05 	& 27 days	& 16172		& 1.8M	   & 0.95M\cr
\TTHREE\	& PoP 2 	& 20/03-30/04 	& 40 days	& 17242		& 2.4M	   & 1.24M\cr
\TFOUR\		& Pop 3   	& 20/03-25/04 	& 35 days	& 31124		& 3.8M	   & 1.76M\cr
\hline
\end{tabular}
\label{tab:desc-traces}
\vspace{-3mm}
\end{table}

In this work, we rely on passive measurements to characterise the
popularity profiles of \youtube videos in operational networks. We
employed Tstat~\cite{finamore2011_tstat}, an open-source traffic
monitoring tool developed at Politecnico di Torino, to analyse the
packets exchanged by end-users from monitored vantage points.  Tstat
was installed on three PoPs (located in different large cities) of a
large Italian ISP, connecting residential customers through ADSL and
FTTH access technologies.  Measurements were collected on both
incoming and outgoing traffic carrying \youtube videos for a period of
three months, from mid March 2012 to late May 2012. During this
period, we observed the activity of more than 60,000 end-users
accessing the Internet normally.  The traces consider only TCP flows
corresponding to \youtube downloads. In total, we recorded almost 10
million transactions, accounting for approximately 227 TB of delivered
content.  Table~\ref{tab:desc-dataset} provides more details on the
traces considered in our analysis; note that here, each IP is
associated to one household. We verified that findings from \TALL\ are
general and not biased, since the results of their analysis are all
almost identical.

%% file: analysis3.tex

\section{Towards a New Model}
\label{sec:analysis}

\tgifeps{7}{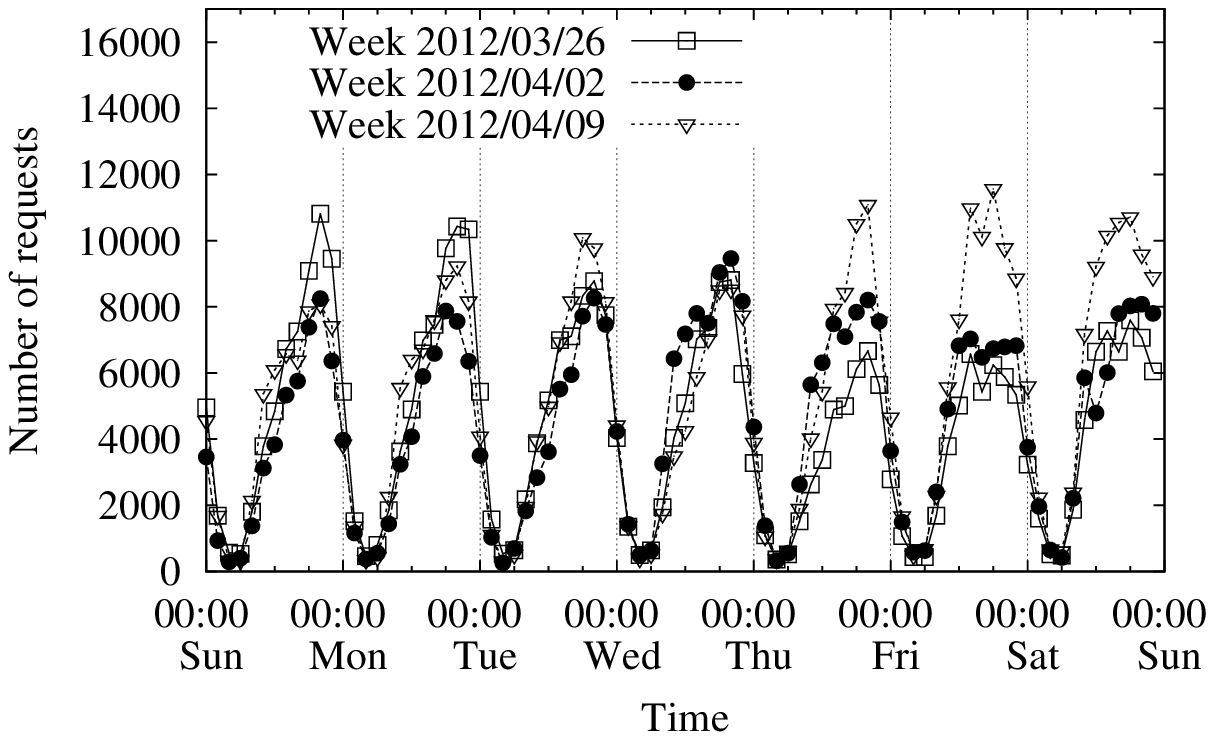}{Evolution of the
  volume of requests over three weeks for \TTHREE. Requests here are
  binned in two-hour windows over the week.}

\tgifeps{7}{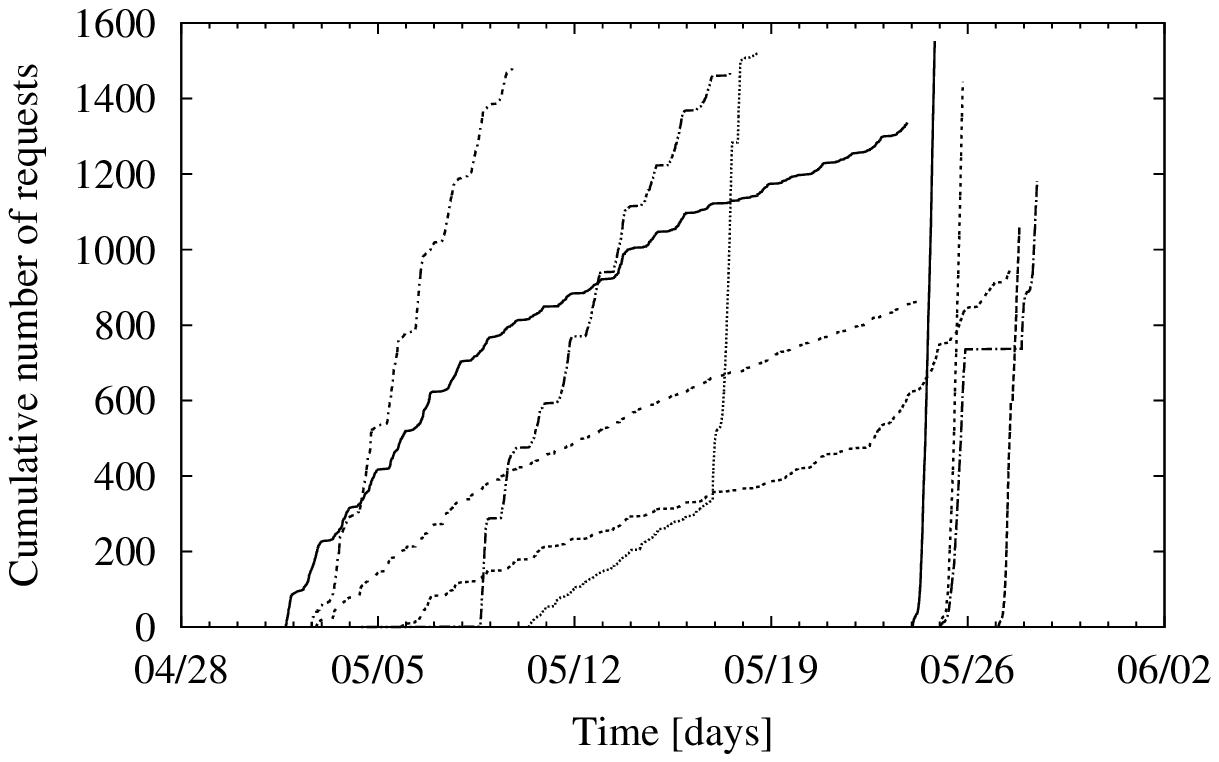}{Cumulative number of
  requests over time for a subset of videos in {\TTWO}; only the
  requests within the life-span of the content are shown..}

\tgifeps{8}{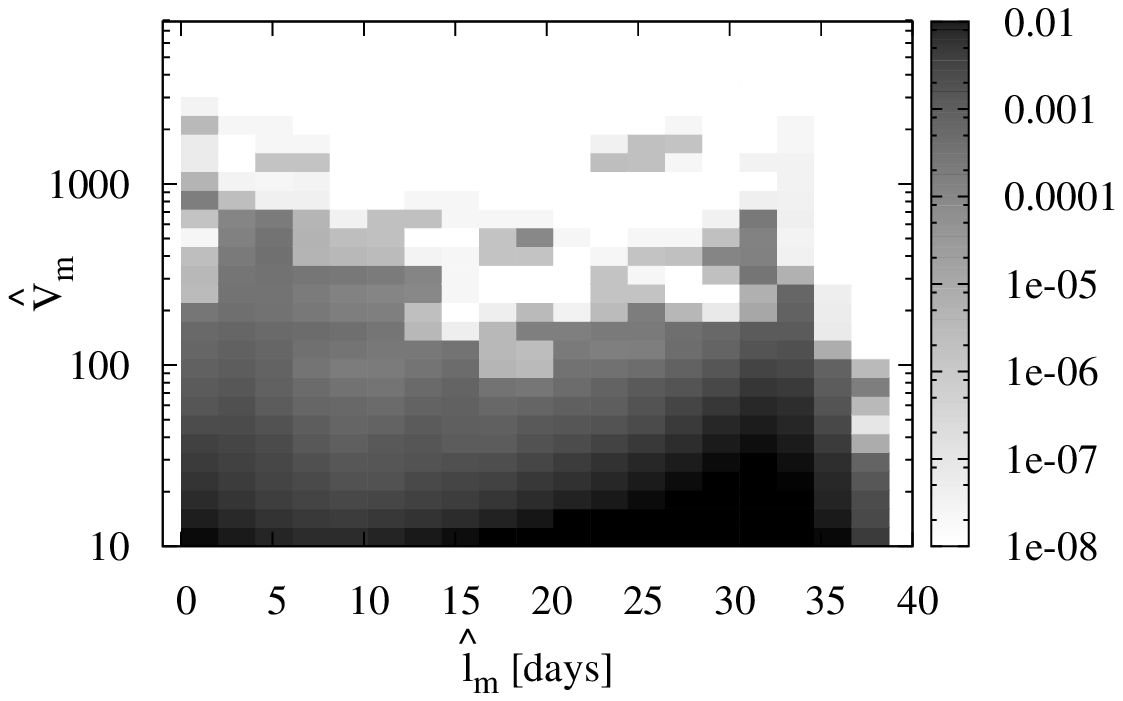}{Density map of contents with
  $\hat{V}_m \ge 10$ observed in \TTHREE, based on estimated number of
  requests $\hat{V}_m$ and effective life-span $\hat{l}_m$. \vspace*{-3mm}}

In this section we study the characteristics of the traffic traces
with the goal of identifying the main factors that should be
considered in order to describe the content request process.

There are two main factors responsible for the non-stationarity
observed in real traffic: (i) the typical diurnal variation of the
aggregate arrival rate of requests (see
Fig.~\ref{fig:reqs_vs_time_week_profile_trace_03}), and (ii) the fact
that the arrival rate of requests for individual contents is highly
non-stationary (see Fig.~\ref{fig:reqs_vs_lifetime_trace_02}).  We find
that, contrary to common opinion~\cite{abrahamsson_imc2012}, the
diurnal variation has little impact on the main performance metrics
for caches (e.g., hit probability, see Sec.~\ref{sec:mv}).  To
understand why this is the case, consider that the hit probability of
almost all proposed caching policies depends only on the sequence of
content IDs arriving at the cache, and not on the time-stamps
associated with the requests. In other words, if we were to
arbitrarily densify or dilute the sequence of content IDs over time,
we would obtain the same hit probability.

With respect to the non-stationarity of content popularity, from
Fig.~\ref{fig:reqs_vs_lifetime_trace_02}, we see that \youtube videos
display extremely heterogeneous request distributions and exhibit
strong time-localities.  For instance, we observe that the popularity
of some videos vanishes after only a few days, while others continue
to attract requests for almost the entire duration of the trace -
reflecting the diversity in user interest. As a result, to capture
the evolution of content popularity over time, we focus just on the
this cause, and characterise each individual content object $m$ with
the following two parameters:

\begin{itemize}[leftmargin=*,nolistsep,noitemsep,topsep=0pt,parsep=0pt,partopsep=0pt]
\item The total number of requests ($V_m$) generated by the content.
\item The effective life-span ($l_m$) of the content, which is defined
  as the duration of the interval in which we see the bulk of its
  requests\footnote{More precisely, the effective life-span ($l_m$) of
    an object is defined as the time elapsing between the two requests
    corresponding to $0.1 V_m$ and to $0.9 V_m$, respectively. The
    effective life-span permits us to filter out the impact of
    outliers on the average lifetime of an object (e.g., one more
    request arriving a long time after the previous requests).}.
\end{itemize}

However, because in practice we only see the requests arriving within
a finite time window (i.e., the length of the trace), both of these
quantities cannot be evaluated exactly from on-the-fly
observations. We will therefore denote by $\hat{V}_m$ and $\hat{l}_m$
the estimated values of $V_m$ and $l_m$, respectively, obtained by
considering only the requests appearing in the trace.  We note that
both variables tend to be underestimated with respect to their actual
values, especially for contents whose true life-span is comparable to
or greater than the trace length.

The density map in Fig.~\ref{fig:density_map_trace_03} reveals that, as
expected, contents exhibit wide heterogeneity in terms of estimated
life-spans $\hat{l}_m$ and estimated volumes $\hat{V}_m$.  The map
also shows that there exists a peculiar correlations between
$\hat{l}_m$ and $\hat{V}_m$. This suggests that a traffic model should
consider the joint distribution of these metrics. In fact, from the
results, we observe that a non-marginal share of videos ($7 - 10\%$)
exhibit a very small life-span ($\hat{l}_m \le 5$ days), while $2\%$
of videos have $\hat{V}_m \ge 10$, but account for a share of requests
that is greater than $27\%$ (these results hold for all the traces in
our data set). These two observations show that a precise traffic
generator should accurately model videos with short life-span, while
accounting for the largest share in requests generated.

At this point, it is worth emphasising that the two parameters $V_m$
and $l_m$ alone do not completely characterise the temporal evolution
of content popularity, which as shown
in Fig.~\ref{fig:reqs_vs_lifetime_trace_02}, can exhibit complex
growth patterns. In fact, recent studies~\cite{Yang:2011, mat2012,
  Ahmed:2013} conducted on much larger data sets reveal that the
popularity of different contents, including videos, follow a limited
number of ``archetypal'' temporal profiles, which essentially depend
on the nature of the content and on the way it becomes
popular. However, as we will see in Sec.~\ref{sec:mv}, cache
performance is essentially driven by the parameters $V_m$ and $l_m$,
while the shape of the popularity profile has only a second-order
effect.  Therefore, a more accurate representation of popularity
evolution, in terms of detailed temporal profiles, is not strictly
necessary in order to accurately predict cache performance.  The
practical side effect of this observation is that we can design a
simple, accurate and robust model to reason about the temporal
evolution of content popularity.


%% file: traffic2.tex

\section{Shot noise traffic model}
\label{sec:traffic_model}
Given the observations in the previous section, we now turn to
proposing a novel approach to describing the arrival process of
content requests generated by a large population of users.  The goal
here is to maintain the generality and flexibility of the IRM
approach, but improve on its accuracy without significantly increasing
the model's complexity. In more detail, the model must: (1) be general
and flexible; (2) provide a native explanation for the temporal
locality of the request process; (3) explicitly represent content
popularity dynamics; (4) capture the phenomena that have a major
impact on cache performance while ignoring those with no or limited
impact; (5) be as simple as possible while maintaining accuracy; (6)
enable users to analytically investigate the performance of popular
caching policies.

Our proposed solution is to represent the overall request process as
the superposition of many independent processes, each referring to an
individual content. In particular, each content $m$ is characterised
by three physical parameters ($\tau_m$, $V_m$, $\lambda_m(t)$):
$\tau_m$ represents the time instant at which the content
enters the system (i.e., when it can be requested by the users); $V_m$
denotes the (average) number of requests generated by the content; and
$\lambda_m(t)$ is the ``popularity profile'', describing how the
request rate for a given object $m$ evolves over time.  In general,
$\lambda_m(t)$ is defined to be a function satisfying the following
conditions: (positiveness) $\lambda_m(t) \ge 0$, $\forall t$;
(causality) $\lambda_m(t)=0$, $\forall t<0$; 
(integrability and normalisation) $\int_0^\infty \lambda_m(t) \diff t
=1$.

Given the above parameters, our model assumes that the request process
for a given content $m$ is described by a time-inhomogeneous Poisson
process whose instantaneous rate at time $t$ is given by:
\vspace*{-1mm}
\[
V_m \lambda_m(t- \tau_m) 
\]

For the sake of simplicity, we assume that new contents become
available in the system according to a homogeneous Poisson process of
rate $\gamma$, i.e., time instants $\{\tau_m\}_m$ form a standard
Poisson process.  We refer to this model as Shot Noise Model (SNM),
since the overall process of requests arrival is known as a Poisson
shot-noise process~\cite{shot}. Fig.~\ref{fig:non-stationary}
illustrates an example of the request pattern generated by the
superposition of two ``shots'' corresponding to two contents with
quite different parameters.

\tgifeps{8}{}{Example of requests (denoted by arrows)
  generated by two contents with different catalogue insertion times
  ($\tau_1, \tau_2$), average number of requests ($V_1,V_2$) and
  profiles ($\lambda_1(t)$,$\lambda_2(t)$).}

We emphasise that the above Poisson assumptions on the instantaneous
generation process of requests for each content, and on the arrival
process of new contents, are essentially introduced for the sake of
analytical tractability.  However, they are very well justified by the
experience gained from our traces.  In fact, the results in
Fig.~\ref{fig:phit_vs_cachesize_shuffle_flip_trace_01} suggest that we
need not be very concerned with temporal correlations in very short
time-scales (in the order of a few hours, say less than 6): removing
all correlations at very short time scales (as given in
Fig.~\ref{fig:phit_vs_cachesize_shuffle_flip_trace_01}) has no
significant impact on the resulting cache performance.  The practical
consequence is that we do not need to take into account possibly
complex correlations in the arrival process of requests such as might
be induced by popularity
cascades~\cite{mat2012,crane2008,Cha:2009}. While cascades in
popularity are indeed observed in large catalogues with a
geographically distributed user base, this is not very evident in our
traffic traces, which are much more local.  Given that short time-scale
correlations (up to a few hours) are not important (in contrast to the
scenarios considered in~\cite{mat2012,crane2008}), the adoption of a
Poisson model for the arrival process of requests is well justified,
and enables us to build simple analytic models for cache performance
analysis such as those developed in~\cite{conext}.

For each given content, the SNM requires that users specify its entire
popularity profile in the form of the function $\lambda_m(t)$, which,
given the difficulty in estimating popularity profiles from a trace,
could be considered as a limitation.  However, we have found that it
is not necessary to precisely identify the shape of $\lambda_m(t)$. In
fact, a simple first-order approximation, according to which we just
specify the content life-span $l_m$,
is enough to obtain accurate predictions of the performance of the
cache. In other words, we can arbitrarily choose any reasonable
function $\lambda_m(t)$ with an assigned life-span $l_m$, and obtain
almost the same results in terms of cache performance (see
 Fig.~\ref{fig:phit_vs_cachesize_fertility_flip_trace_04_paolo}).
Finally, content heterogeneity is taken into account by associating a
life-span $l_m$ with every content, jointly with the (typically
correlated) total number of requests $V_m$.  This means that, upon
arrival of each new content $m$, independently for each content, we
randomly choose the pair of parameters ($V_m$, $l_m$) from a given
assigned joint distribution.

\subsection{Parameter Fitting}\label{subsec:fit}
There are many ways to derive the parameters of the SNM from a trace.
In this section, we present the simple approach which we have employed
to fit our data traces.  We first partition the contents into $6$
classes ($0,\ldots,5$), on the basis of the measured request volume
$\hat{V}_m$ and content life-span $\hat{l}_m$.  Content Class $0$ is
defined to contain objects with $\hat{V}_m < 10$, representing
contents for which we cannot derive a reliable estimate of the
life-span, because they gather too few requests. Due to this, we treat
contents in this class using the IRM approach, and assume that their
requests fall uniformly at random within the synthesised trace.  It is
worth acknowledging that, in doing this, we lose an opportunity to
characterise the time locality of a significant fraction of contents
($85\%$ of the contents have $\hat{V}_m < 10$).  However, even with
this restriction, as we will see, we can still obtain a reasonable
conservative prediction on the resulting cache performance, (i.e., we
slightly overestimate the cache size required to achieve a desired hit
ratio).

\input{bigtable}

Classes $1$ to $5$ contain contents with $\hat{V}_m \geq 10$ and, as
shown in Table~\ref{tab:desc-dataset}, are partitioned based on their
estimated life-spans ($\hat{l}_m$). For each content class,
Table~\ref{tab:desc-dataset} reports the percentage of total requests
attracted by the class, the percentage of contents belonging to it,
and their average estimated values $\hat{V}_m$ and $\hat{l}_m$.
Notice from Table~\ref{tab:desc-dataset} that contents in Class 1, whose
life-span is smaller than $2$ days, represent less than $4\%$ of the
total number of contents but attract approximately $10\%$ of all
requests. Therefore, because these contents exhibit a large degree of
temporal locality, they can be expected to have significant impact on
cache performance.

Observe also from Table~\ref{tab:desc-dataset} that the values related
to each class are quite similar across the four traces (within a
factor of $2$). This is significant, because it suggests that our
broad classification captures some invariant properties of the
considered traffic. Therefore, we have the opportunity of building a
synthetic traffic model that is fairly general and flexible, by
fitting the parameters that appear to be invariant (or almost
invariant) from one single trace, and properly scaling those
parameters that clearly depends on the particular context (such as the
number of end-users).

However, because our traces cover a period of approximately one month
each, they are not long enough to extract very general laws,
especially for long-lived contents. This is particularly evident for
contents in class $5$ ($\hat{l}_m > 13$ days), whose life-span is
comparable with the trace length.  Therefore, their estimated value of
$\hat{l}_m$ is expected to be strongly affected (i.e., underestimated)
by border effects due to the trace being finite. For this reason, to
obtain a conservative prediction, we again treat contents in this
class as if they had a stationary popularity (like in the IRM),
generating their requests uniformly in the considered time horizon.

In sum, given the six broad classes of behaviour defined, only objects
in Classes $1$ to $4$ (see Table~\ref{tab:desc-dataset}) are generated
according to the SNM model as follows.  First, based on the estimated
rate at which contents are generated in each class (see column
``Videos" in Table~\ref{tab:desc-dataset}), we sample the time instants
$\{\tau_m\}$ at which contents of the class become available. This can
be done in a standard way, because the arrival process of contents
within each class is assumed to be homogeneous Poisson. Second, we set
the life-span of all the contents of the class equal to the
corresponding estimated average life-span $E[\hat{l}_m]$, thereby
compensating for border effects due to content life-spans comparable
with the trace duration.  Third, we randomly choose $V_m$ for each
content generated in the class according to the corresponding
empirical distribution observed in the trace.  Finally, the events
corresponding to content requests are generated for each content $m$
according to an inhomogeneous Poisson process with fertility function
$V_m\lambda(t-\tau_m)$ according to standard methods~\cite{ross06}.
In Sec.~\ref{sec:compu}, we will discuss the computational cost
associated with the SNM.

Moreover, for objects in Classes $1$ to $4$ we have chosen a unique
``shape'' ($\lambda_{m}(t)$) for the popularity profile of all
contents and considered two different shapes, with parameter $L$
(content lifetime) as: (1) exponential popularity $\lambda(t)=
\frac{1}{L}e^{-\frac{t}{L}}$ for $t\geq 0$; (2) uniform popularity
$\lambda(t)= \frac{1}{2L}$ for $t\in[0,2 L]$.  Note that, for each
content class, $L$ is chosen to match the desired life-span
$E[\hat{l}_m]$~\footnote{For instance, in the case of the uniform
  shape, $L = \frac{0.5 E[\hat{l}_m]}{0.8}$.}.

\tgifeps{8}{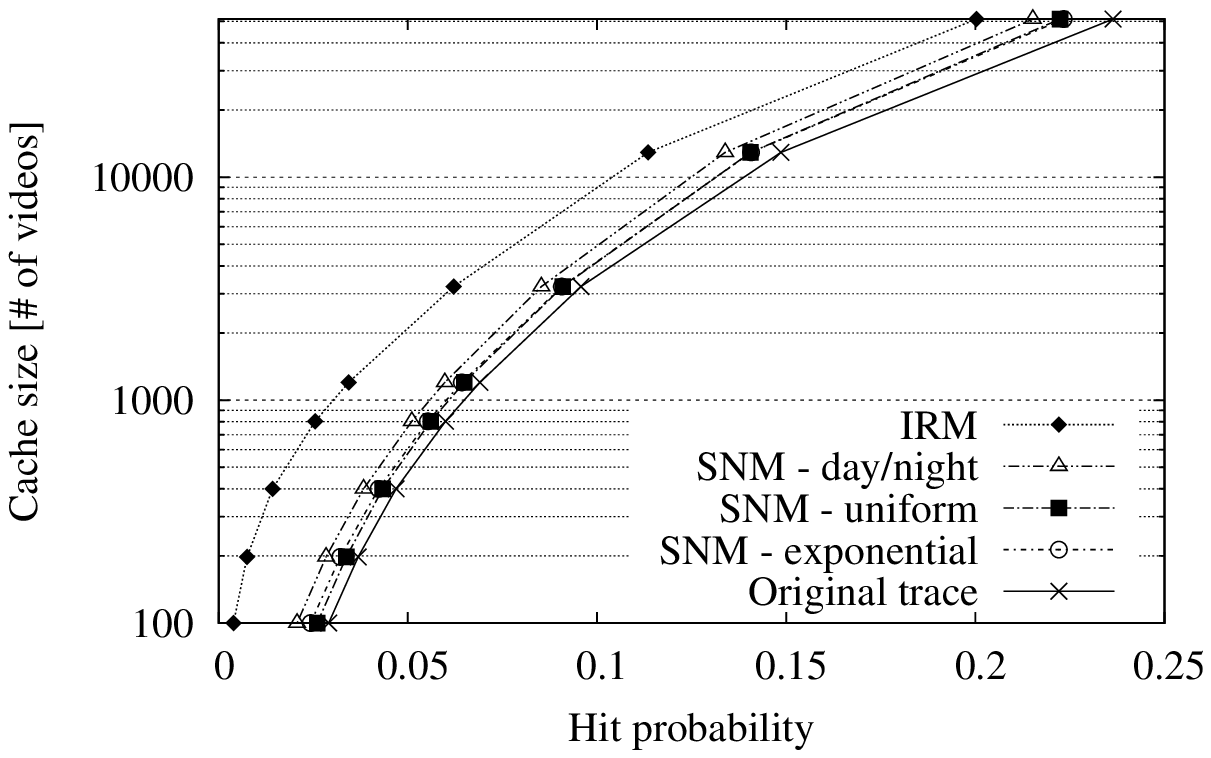}{Cache
  size vs hit probability under LRU policy for \TFOUR. Note that
  similar results are obtained for all traces.}
  
\subsection{Model Validation}\label{sec:mv}
To evaluate the accuracy of our traffic model we generate synthetic
request traces as described in Sec.~\ref{subsec:fit} and feed them to
a cache implementing the LRU policy. We again report the cache size
required to achieve a desired hit probability.  For comparison, we
also report the results obtained with the original trace and its
completely shuffled version. The latter neglects the contents'
temporal locality and provides the same performance as the IRM
approach; based on long-term measurements of the content popularity
distribution.

From Fig.~\ref{fig:phit_vs_cachesize_fertility_flip_trace_04_paolo} we
find that the results obtained using the SNM (using either the uniform
or the exponential shape) are very close to those obtained with the
original trace. As mentioned, the shape of the popularity profile
chosen (exponential/uniform) has little impact on the results.  While
in contrast, the discrepancies produced by the shuffled trace (IRM)
are quite large, especially for small caches.  This confirms that the
impact, on cache performance, of a small number of highly popular
contents with a relatively short life-span (in the order of a few
days) is significant and should not be neglected.

The results
in Fig.~\ref{fig:phit_vs_cachesize_fertility_flip_trace_04_paolo} show
that the SNM provides an accurate prediction of cache performance,
despite the heavy simplifications adopted in the parameter
identification. We expect that even better predictions could be
achieved by improving the fitting procedure, or, if available using
much longer traces.

Finally, Fig.~\ref{fig:phit_vs_cachesize_fertility_flip_trace_04_paolo}
also reports (label Day/Night) an extended version of the model that
incorporates the effects of daily oscillations in the traffic
rate. This is produced by simply modulating the shots of all contents
by a fixed, periodic function of time $f(t)=1+ \sin( 2 \pi t)$ (where
$t$ is expressed in days).  From this result, we see that the daily
variations have a marginal impact on the cache
performance.\footnote{The adopted modulation technique does not
  exactly produce the effect of just stretching/squeezing the sequence
  of requests generated by the un-modulated process, which explains
  why the impact on cache performance is not null.}

\subsection{Computational cost}\label{sec:compu}
Due to its simplicity, the SNM can be effectively employed to generate
synthetic traces for large-scale simulations, permitting users to
explore realistic scenarios that could hardly be studied by employing
experimental traces alone.  More precisely, the computational cost to
generate a synthetic trace using the SNM is only slightly larger than
the cost for the IRM. When a new content with parameters $V_m$ and
$l_m$ is introduced to the system, we can first generate the total
number of requests received by this content according to a Poisson
distribution of mean $V_m$, then schedule ahead of time; the arrival
time of each request according to the content's temporal profile
(which depends on $l_m$), recording it into an ordered or partially
ordered data structure (e.g., a heap), to be dynamically used in an
event-driven simulator.

Therefore, with respect to the $\Theta(1)$ cost needed to generate a
request under the IRM, the cost to generate a request under the SNM
equals the insertion time of an element into an ordered data
structure; which is $\Theta(\log{M})$, where $M$ is the number of
requests scheduled in the future.  Notice that, whilst in the worst
case we can take $M$ equal to the trace length. In practice $M$ is
much smaller than this, since it is related to the average number of
requests falling in a shot, i.e., $M =
\Theta(\gamma\,\mathbb{E}(V_m\,l_m))$.  From our experience, the
additional time needed to simulate a cache under our synthetic traffic
model, compared to the time needed when an experimental trace is
available, is marginal.

\section{Final Considerations}
This work, to the best of our knowledge, presents the first
quantification of the error arising from applying the IRM to the
evaluation of cache performance in the context of on-demand video
delivery.  We have shown that the error that the IRM induces is large
enough to render it too pessimistic for practical use, especially in
scenarios with small cache sizes. As a replacement, we proposed a
parsimonious and novel approach termed the Shot-Noise Model (SNM) that
accurately models cache performance while being sufficiently simple to
be effectively employed in both analytical and scalable simulative
studies of large caching systems.

The SNM provides a flexible and accurate approach to model and
synthesise contents' request arrival process and does so by capturing
some of the fundamental characteristics of temporal locality. This
paper provides an accurate tool to investigate and understand the
performance of caches in content-driven systems. Having such a tool
enables the development of accurate forecasts of the network's
performance, leading to a more efficient dimensioning of content
delivery systems and ultimately reducing operating costs while
improving the experience of end-users.
 
\section{Acknowledgments}
We thank Marco Mellia, Felipe Huici, Alessandro Finamore and the
anonymous reviewers for providing useful feedback on earlier drafts of
this paper. This research is supported by funding from the European
Union under the FP7 Grant Agreement no. 318627 (Integrated Project
``mPlane'').

%% file: bigtable.tex
\begin{table}[t!]
\centering
\scriptsize
\caption{Model parameters for content classes 1--5.}
\begin{tabular}{|c|c|c|c|c|c|c|}
\hline
Class	& Life-span [days]		& Trace 	& \%Reqs& \%Videos	& $E[\hat{l}_m]$ 	& $E[\hat{V}_m] $\cr\hline 
\multirow{4}{*}{Class 1}	& \multirow{4}{*}{$\hat{l} \le 2$} 	&\TONE\ 	& 9.15	&3.17 	  	& 1.14 		& 86.4	       	\cr
				&	&\TTWO\		& 10.05 & 4.17 		& 1.09 		& 76.2       	\cr
				&	&\TTHREE\	& 9.44 	& 3.73 		& 1.04 		& 76.0       	\cr
				&	&\TFOUR\	& 7.77 	& 3.34 		& 1.06 		& 74.0       	\cr\hline
\multirow{4}{*}{Class 2}	& \multirow{4}{*}{$2<\hat{l}\le 5$} 	&\TONE\ 	& 6.80 	& 4.9  	  	& 3.36 		& 41.9        	\cr
				&	&\TTWO\		& 12.55	& 7.83 		& 3.34 		& 50.7       	\cr
				&	&\TTHREE\	& 6.55 	& 4.54 		& 3.32 		& 43.3       	\cr
				&	&\TFOUR\	& 6.12 	& 4.06 		& 3.41 		& 48.0       	\cr\hline
\multirow{4}{*}{Class 3}	& \multirow{4}{*}{$5<\hat{l}\le 8$} 	&\TONE\ 	& 5.87 	& 2.95 		& 6.40 		& 59.5		\cr
				&	&\TTWO\		& 6.72 	& 4.74 		& 6.31 		& 44.9	\cr
				&	&\TTHREE\	& 6.05 	& 2.87 		& 6.42 		& 63.3	\cr
				&	&\TFOUR\	& 5.14 	& 2.71 		& 6.45 		& 60.3	\cr\hline
\multirow{4}{*}{Class 4}	& \multirow{4}{*}{$8<\hat{l}\le 13$}	&\TONE\ 	& 5.49 	& 4.45 		& 10.53 	& 36.9         \cr
				&	&\TTWO\		& 10.79 & 8.61 		& 10.86 	& 39.6        \cr
				&	&\TTHREE\	& 4.84 	& 3.68 		& 10.62 	& 39.5        \cr
				&	&\TFOUR\	& 5.34 	& 4.48 		& 10.65 	& 37.8        \cr\hline
\multirow{4}{*}{Class 5}	& \multirow{4}{*}{$\hat{l} > 13$}	&\TONE\ 	& 72.69 & 84.58		& 24.61 	& 25.7		\cr
				&	&\TTWO\		& 59.89 & 74.65 	& 19.29 	& 25.3	\cr
				&	&\TTHREE\	& 73.11 & 85.17 	& 28.19 	& 25.8	\cr
				&	&\TFOUR\	& 75.63 & 85.41 	& 24.59 	& 28.1	\cr\hline
\end{tabular}
\label{tab:desc-dataset}
\vspace{-3mm}
\end{table}